\documentstyle[sprocl]{article}

\input{psfig}

\def\be{\begin{equation}}
\def\ee{\end{equation}}
\def\bea{\begin{eqnarray}}
\def\eea{\end{eqnarray}}

\begin{document}

\title{MISSING NUCLEON RESONANCES IN KAON PRODUCTION WITH PIONS AND PHOTONS}

\author{C. BENNHOLD, A. WALUYO. H. HABERZETTL}

\address{Center for Nuclear Studies, Department of Physics, The George 
         Washington University, \\Washington, D.C. 20052, USA\\E-mail:bennhold@gwu.edu}

\author{T. MART}

\address{Jurusan Fisika, FMIPA, Universitas Indonesia,\\ Depok 16424, Indonesia} 

\author{G. PENNER, U. MOSEL}

\address{Institut f\"ur Theoretische Physik, Universit\"at Giessen,\\
         D-35392 Giessen, Germany}

\maketitle\abstracts{
Resonances with strong couplings to strangeness channels are investigated
in a coupled-channels framework of kaon production with pions and photons. 
We find evidence for an additional $D_{13}$ resonance around 1900 MeV that has a significant
decay width into the $K \Lambda$ final state.  
Such a state has been predicted in a relativistic quark model 
at 1960 MeV with significant $\gamma N$ and $K \Lambda$ branching ratios.}

\section{Introduction}

Like any composite system the nucleon can be excited into a number of resonant
or compound states. However, due to the nonperturbative nature of
QCD at nucleon resonance energies no rigorous understanding of $N^*$ resonances
on the basis of QCD has emerged. 
Lattice QCD has only recently begun to contribute
to this field.  In a recent study \cite{lee98} the excitation energies of 
$1/2^-$ and $3/2^-$ baryon resonances are calculated for the first time 
on the lattice with improved actions.  On the other hand, 
quark models that contain three constituent valence quarks
predict a much richer resonance spectrum \cite{NRQM,capstick94}
than has been observed in $\pi N\to \pi N$ scattering experiments. 
Quark model studies have suggested that those "missing" resonances 
may couple strongly to other channels, 
such as the $K \Lambda$ and $K \Sigma$ channels \cite{capstick98}
or final states involving vector mesons.  
While most of the present information on nucleon resonances has been
obtained in $\pi N\to \pi N$ experiments the advent of the
newly established electron and photon facilities have made 
it possible to investigate the mechanism of nucleon resonance excitation
with photons and electrons. 
Experiments producing kaon-hyperon final states have been performed at 
{\small ELSA} \cite{saphir98} and are being analyzed at JLab.

\section{Kaon Photoproduction at Tree Level}

Within the one-photon approximation, the full amplitude for any photoproduction process can be 
written in terms of a Bethe-Salpeter equation
\begin{eqnarray}
M ~ = ~V ~+ ~ V~G~T,
\label{eq:tmatrix}
\end{eqnarray} 
where $V$ represents the driving term for the particular photoproduction
process, $G$ is the meson-baryon two-particle propagator, and $T$ is the
hadronic meson-baryon final state interaction.  In principle, one would
have to solve this equation as a four-dimensional integral equation. However,
most studies over the last 30 years 
have analyzed the $N(\gamma, K)\Lambda (\Sigma)$ reactions in a tree-level isobar 
framework \cite{adelseck85}, neglecting
the hadronic final state interaction in Eq. 1.
Without rescattering contributions the $T$-matrix is simply
approximated by the driving term alone which is assumed to be given by 
a series of tree-level diagrams whose couplings were adjusted to
reproduce the experimental data.  Due to the poor data quality it was 
not possible to decide which resonances contributed, even the 
background terms could not be modelled properly, i.e., if standard SU(3)
coupling constants were used for the leading $g_{K \Lambda N}$ and $g_{K \Sigma N}$
couplings the calculations overpredicted the data by an order of magnitude.
The recent work on including hadronic form 
factors in photoproduction reactions \cite{hbmf98} while maintaining 
gauge invariance has resulted in the proper description
of the background terms, allowing the use of approximate SU(3) symmetry to 
fix the Born coupling constants $g_{K \Lambda N}$ and $g_{K \Sigma N}$.

Here, we use as a starting point
the tree-level isobar model described in Ref.\,\cite{elba98,mart99} to analyze the
$p(\gamma, K^+)\Lambda$ process in more detail.  
Guided by a recent coupled-channels analysis 
\cite{feuster98}, the low-energy resonance part of
this model includes three states that have been found to have 
significant decay widths into the $K^+\Lambda$ channel, 
the $S_{11}$(1650), $P_{11}$(1710), and $P_{13}(1720)$ resonances. 
The background part includes 
the standard Born terms along with the $K^*$(892) and $K_1$(1270) 
vector meson poles in the $t$-channel. As in Ref.\,\cite{elba98}, we 
employ the gauge method of Haberzettl \cite{hbmf98}
to include hadronic form factors.

\begin{figure}[!ht]
\centerline{\psfig{file=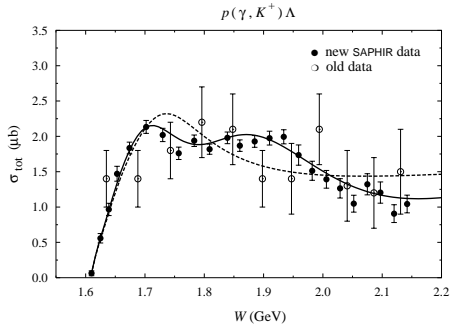,height=4.5cm,clip=,silent=}}
\vspace{-0.2cm}
\caption[thanks]{\label{fig:totgamktree}
        Total cross section calculations for $K^+\Lambda$ photoproduction
        on the proton obtained at tree level. The dashed line shows the result without the 
        new $D_{13}(1900)$ resonance, while the solid line is obtained 
        by including the new $D_{13}(1900)$ state. The new {\small SAPHIR} 
        data \protect\cite{saphir98} are denoted by the solid 
        circles.}        
\end{figure}

The new {\small SAPHIR} total cross section data \cite{saphir98} for 
the $p(\gamma, K^+)\Lambda$ channel, shown in Fig.~\ref{fig:totgamktree},
indicate for the first time a 
structure around $W = 1900$ MeV. This structure could not be resolved before 
due to the low quality of the old data. While there are no 3 - or 4-star 
isospin 1/2 resonances around 1900 MeV in the 
Particle Data Book \cite{pdg98}, several 
2-star states are listed. Of those only the
$D_{13}(2080)$ has been identified in older $p(\pi^-, K^0)\Lambda$ 
analyses \cite{saxon,bell} to have a noticeable branching ratio into 
the $K \Lambda$ channel.  On the theoretical side, the constituent quark 
model by Capstick and Roberts \cite{capstick94} predicts many new states 
around 1900 MeV, however, only a few them have been calculated to have
a significant $K \Lambda$ decay width \cite{capstick98}.
These are the $[S_{11}]_3$(1945), $[P_{11}]_5$(1975), 
$[P_{13}]_4$(1950), and $[D_{13}]_3$(1960) states,
 where the subscript refers to the 
band that the state is predicted in. As discussed in more detail
in ref. \cite{mart99} we have performed fits for each 
of these possible states, allowing the fit to determine the mass, width 
and coupling constants  of the resonance. 
While all four of the above resonances have 
large decay widths into the $K \Lambda$ channel, only the  $D_{13}$(1960) 
state is predicted to also have significant photocouplings. Table 
\ref{table_cc2} presents the remarkable agreement, up to the sign, between 
the quark model prediction and our extracted results for the $D_{13}$(1960).
For the other three states the partial widths extracted from our fit 
overestimate the quark model results by up to a factor of 30.  
Figure \ref{fig:totgamktree} compares our models with and without the
$D_{13}$(1960) with the {\small SAPHIR} total cross section data\cite{saphir98}.
Our result without this resonance shows only one peak
near threshold, while inclusion of the new resonance leads
to a second peak at $W$ slightly below 1900 MeV,
in agreement with the new {\small SAPHIR} data\cite{saphir98}.
As shown in Ref.\,\cite{mart99}, 
the difference between the two calculations is smaller for 
the differential cross section and the recoil polarization.

\begin{table}[!t]
\caption{Comparison between the results from our tree-level fit to 
   the $p(\gamma, K^+)\Lambda$ data  and those 
  of the quark model (QM), where the QM photocouplings were taken 
  from Ref.\,\protect\cite{capstick92}
  and the $K \Lambda$ decay widths from Ref.\,\protect\cite{capstick98}.}
\renewcommand{\arraystretch}{1.4}
\vspace{0.2cm}
\label{table_cc2}
\begin{center}
\begin{tabular}{llrrr}
\hline
\hline
Missing Resonance & Model & $m_{N^*}$ & $\Gamma_{N^*}$ &
$\sqrt{\Gamma_{N^*N\gamma}\Gamma_{N^*K\Lambda}}/\Gamma_{N^*}$ \\
 & & (MeV) & (MeV) & ($10^{-3}$) \\ [0.5ex]
\hline
$D_{13}$ & Fit & ${1895}$ & ${372}$ &
        ${2.292^{+0.722}_{-0.204}}$ \\
               & QM  & ${1960}$ & ${535}$ &
        ${-2.722\pm 0.729}$ \\ [2ex]
\hline
\end{tabular}
\end{center}
\end{table}

\section{Kaon Production in a Coupled Channels K-matrix Approach}

Going beyond tree level requires inclusion of the rescattering term
in Eq. 1. To simplify the integral equation one can rewrite 
the full Bethe-Salpeter equation in the form 
\begin{eqnarray}
K &=& V + V~{\rm Re}(G_{\rm BS})K\\
T &=& K + iK~{\rm Im}(G_{\rm BS})T ~.
\end{eqnarray}
where $G_{BS}$ is the full propagator.  Any truncation of the first 
equation will still provide a unitary, albeit approximate, solution, as long as 
i Im$(G_{BS}$ correctly describes the discontinuity across the scattering 
cut. Taking the special choice
of placing both intermediate particle on shell
leads to $K=V$ and simple K-matrix (Born) approximation:
\begin{eqnarray}
T &=& \frac{V}{1-iV} ~.
\end{eqnarray}

This procedure still allows for the resonance widths to be generated dynamically, while
the real part of the self-energy is absorbed in an effective resonance mass
that is determined by the fit.
Enforcing unitarity dynamically therefore requires
solving a system of coupled channels with all possible final states. 
The most recent work that employs the $K$-matrix
approximation within an effective Lagrangian framework has been performed
by Feuster and Mosel \cite{feuster98}. 
They extract nucleon resonance parameters by simultaneously analyzing
all available data for reactions involving the initial and final states
$\gamma N, \pi N, \pi \pi N, \eta N$ and $K \Lambda$ up to 
$W = 1.9$ GeV.

\begin{figure}[!ht]
\centerline{\psfig{file=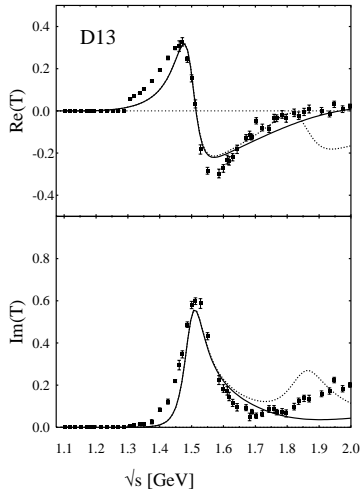,height=6.5cm,clip=,silent=}}
\vspace{-0.1cm}
\caption[thanks]{\label{fig:d13pinphase}
        The $D_{13}$ phase  calculated in the coupled-channels K-matrix
        approach, fitted to the SM95 PWA. The solid line includes the new  
        $D_{13}(1900)$ state while the dashed line does not.}        
\end{figure}

Here we extend the approach of Ref.\,\cite{feuster98} to study 
resonances around 1900 MeV in more detail.  We have extended the energy
range up to 2.0 GeV and included the $K \Sigma$ final state.  Our best fit
to all hadronic data of the reactions $\pi N \to \pi N$, $\pi N \to \pi \pi N$,
$\pi N \to \eta N$, $\pi N \to K \Lambda$ and  $\pi N \to K \Sigma$ indeed does result
in three  $D_{13}$ excitations, the well-known $D_{13}$(1520) state, 
the 3-star $D_{13}$(1700) state and a new $D_{13}$ state at 1945 MeV with a large width
of 583 MeV.  Figure \ref{fig:d13pinphase} shows our description of the $D_{13}$
$\pi N$ partial wave; the prominent $D_{13}$(1520) is clearly visible, the $D_{13}$(1700)
is not apparent, while the inclusion of another state around 1900 MeV clearly improves
the agreement in the real part which exhibits a zero crossing at that energy. 
The imaginary part, on the other hand, is missing some strength, possibly indicating
the threshold opening of a channel that is not included.

\begin{figure}[!ht]
\centerline{\psfig{file=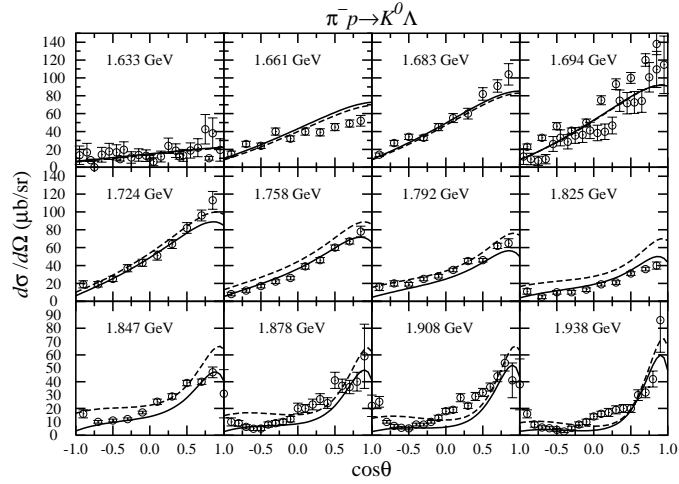,height=6.5cm,clip=,silent=}}
\vspace{-0.2cm}
\caption[thanks]{\label{fig:diffpik}
        Differential cross section calculations for the $p(\pi^-, K^0)\Lambda$ 
        reaction in the coupled-channels approach. The notation of the curves is 
        as in Fig. \ref{fig:d13pinphase}.}        
\end{figure}

\begin{figure}[!ht]
\centerline{\psfig{file=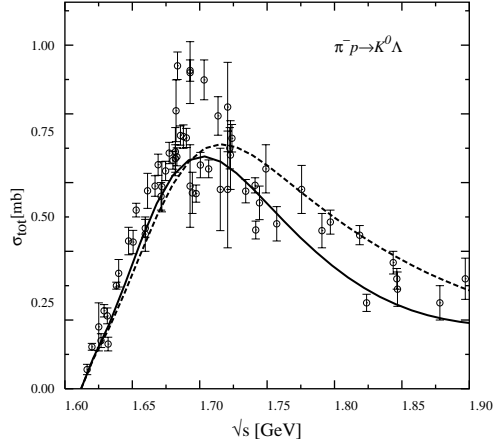,height=6.0cm,clip=,silent=}}
\vspace{-0.2cm}
\caption[thanks]{\label{fig:totpik}
        Total cross section calculations for the $p(\pi^-, K^0)\Lambda$ 
        reaction in the coupled-channels approach. The notation of the curves is 
        as in Fig. \ref{fig:d13pinphase}.}        
\end{figure}

Fig. \ref{fig:diffpik} displays differential cross sections for the 
$\pi^- p \to K^0 \Lambda$ reaction at different W. While the signal is small there is 
a clear improvement in the description of the data when the new $D_{13}$ state is included.
Unfortunately, the total cross section data for the same reaction, 
shown in Fig. \ref{fig:totpik}, are much too poor to distinguish between the two cases.

Is this state identical to the 2-star resonance $D_{13}$(2080) 
listed in the Particle Data Table? Table \ref{table_d13} displays a list of
$D_{13}$ states below 2.1 GeV predicted by 
Refs.\,\cite{capstick94,capstick98}, along with the Particle Data Table
listings and the states found in our fit within the K-matrix approach.
A closer examination of the literature reveals
that there has been some evidence for two resonances in this partial wave
between 1800 and 2200 MeV \cite{cutkosky80}; one with a mass centered around 
1900 MeV and another with mass around 2080 MeV. It is the former
which has been seen prominently in two separate $p(\pi^-, K^0)\Lambda$
analyses \cite{saxon,bell}. 
Thus, even though the remarkable quantitative 
agreement is probably fortuitous, we believe that the structure seen in the
{\small SAPHIR} data is in all likelihood
identical to the one seen in hadronic $K \Lambda$ production
and, furthermore, corresponds to the
$D_{13}$(1960) state predicted by the quark model.
The $D_{13}$ excitation around 2080 MeV seen in Ref.\,\cite{cutkosky80} may 
well correspond to the quark model state $D_{13}$(2055) in the $N=4$ band.

In order to clearly separate these nearby $D_{13}$ states, measuring other
channels will be helpful. For example, Ref.\,\cite{capstick94} predicts the
$D_{13}(1960)$ to have large decay widths into the $\eta N$ and
$\eta 'N$ channels, in contrast to the $D_{13}(2055)$ whose branching 
ratios into these channels are negligible.

\begin{figure}[!ht]
\centerline{\psfig{file=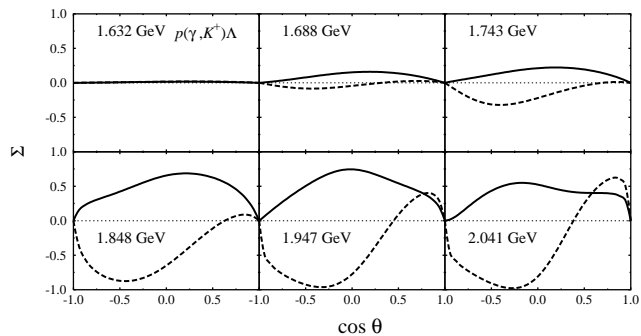,height=4.5cm,clip=,silent=}}
\vspace{-0.2cm}
\caption[thanks]{\label{fig:siggamk}
        Polarized photon asymmetry for the $p(\gamma, K^+)\Lambda$ 
        reaction. The notation of the curves is 
        as in Fig. \ref{fig:d13pinphase}.}        
\end{figure}

What other experiments can help uncover the role played by this resonance?
The largest effects are found in the photon asymmetry 
shown in Fig. \ref{fig:siggamk}. For $W\geq 1800$
MeV, including the new resonance leads to a sign change in the photon 
asymmetry whose magnitude is almost 100 percent at intermediate angles.
Clearly, measuring this observable is 
well suited to shed more light on the contribution of this state 
in kaon photoproduction.
Ultimately, only a detailed multipole analysis can verify that the observed 
structure is indeed due to a resonance.
With the arrival of
new, high-precision cross section and polarization data the kaon 
photoproduction process will be able to 
unfold its full potential in the search and study of nucleon resonances.

\begin{table}[!t]
\caption{Summary of listed $D_{13}$ resonances. The observed states from the 
        Particle Data Table are ordered according to 
        Refs.\,\protect\cite{capstick94,capstick98} and compared to our fit.}
\renewcommand{\arraystretch}{1.4}
\label{table_d13}
\begin{center}
\begin{tabular}{ccc}
\hline
\hline
Quark Model \protect\cite{capstick94,capstick98} & Particle
Data Table \protect\cite{pdg98} & our fit \\ [0.5ex]
\hline
$[N\frac{3}{2}^-]_1 (1495)$ & $D_{13}(1520)$(****) & $D_{13}(1520)$  \\
$[N\frac{3}{2}^-]_2 (1625)$ & $D_{13}(1700)$(***) & $D_{13}(1706)$ \\
$[N\frac{3}{2}^-]_3 (1960)$ & $D_{13}(2080)$(**) & $D_{13}(1945)$ \\
$[N\frac{3}{2}^-]_4 (2055)$ & - &- \\
$[N\frac{3}{2}^-]_5 (2095)$ & - &- \\
[0.5ex]
\hline
\end{tabular}
\end{center}
\end{table}

\section*{Acknowledgments}
This work was supported by US DOE grant DE-FG02-95ER-40907 (CB, AW, and HH),
the University Research for Graduate Education (URGE) grant (TM), and
the DFG and GSI Darmstadt in Germany (GP and UM).

\end{document}